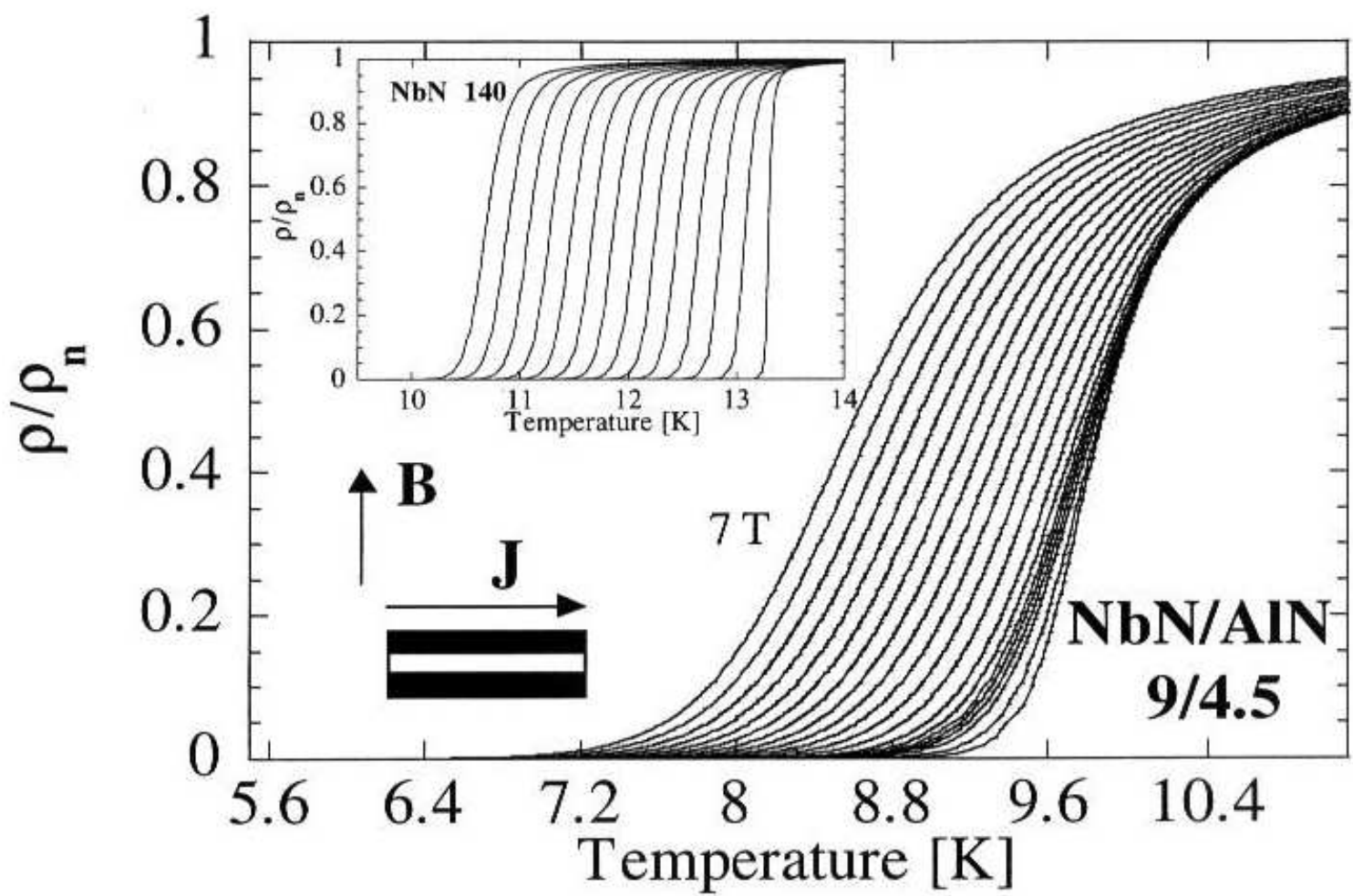

Fig.1. E. S. Sadki *et al*

TABLE I. The derived sample parameters.

| Sample | $T_c$ (K) | $\xi_{ab}(0)$ (nm) | $\xi_c(0)$ (nm) | $\Gamma$ |
|---|---|---|---|---|
| 140 | 13.3 | 3 | ... | ... |
| 9/2.25 | 10.7 | 2.9 | 0.78 | 3.7 |
| 9/4.5 | 9.9 | 2.6 | 0.10 | 26 |
| 9/9 | 9.7 | 2.5 | 0.06 | 42 |

E. S. Sadki *et al*

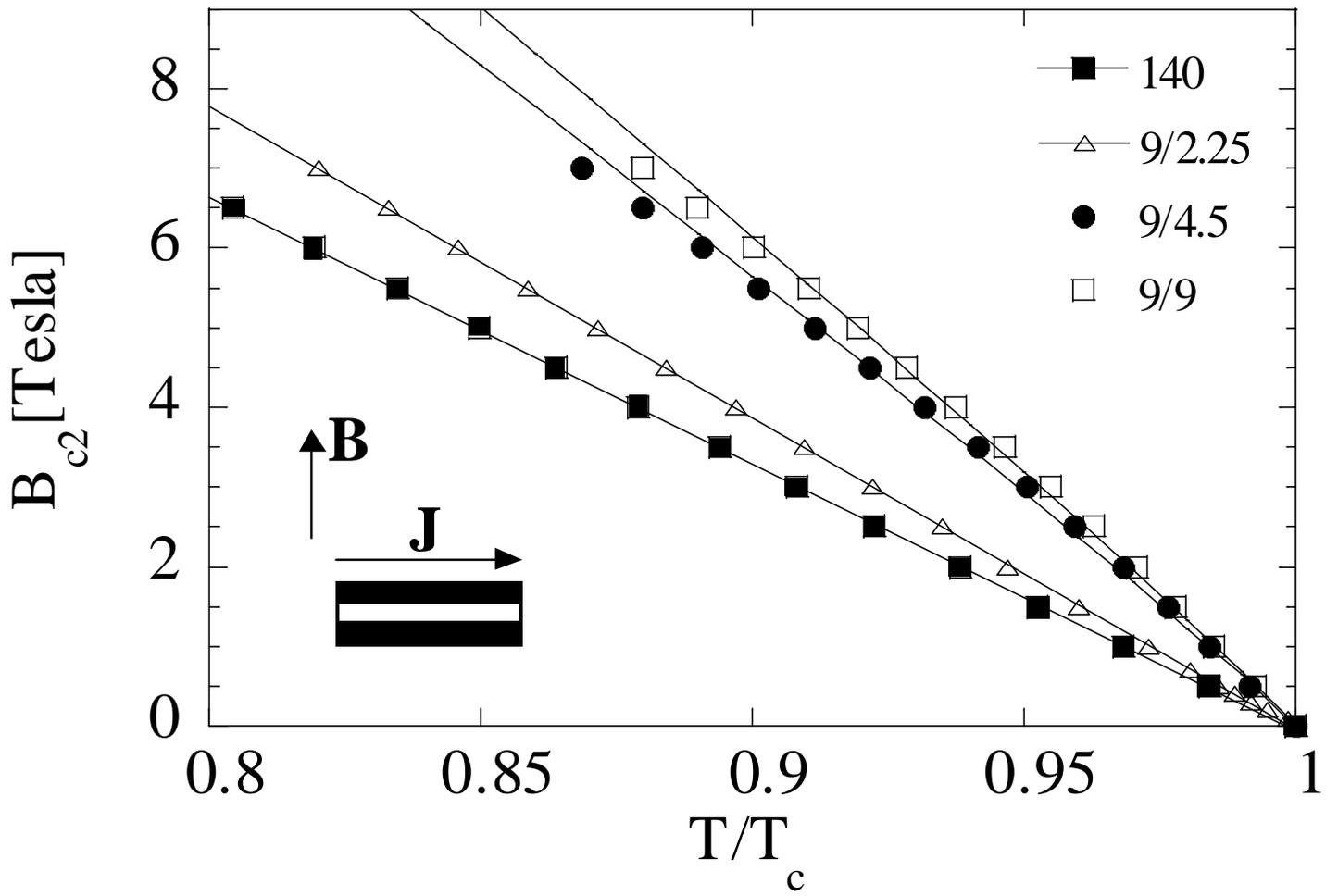

Fig. 2. E. S. Sadki *et al*

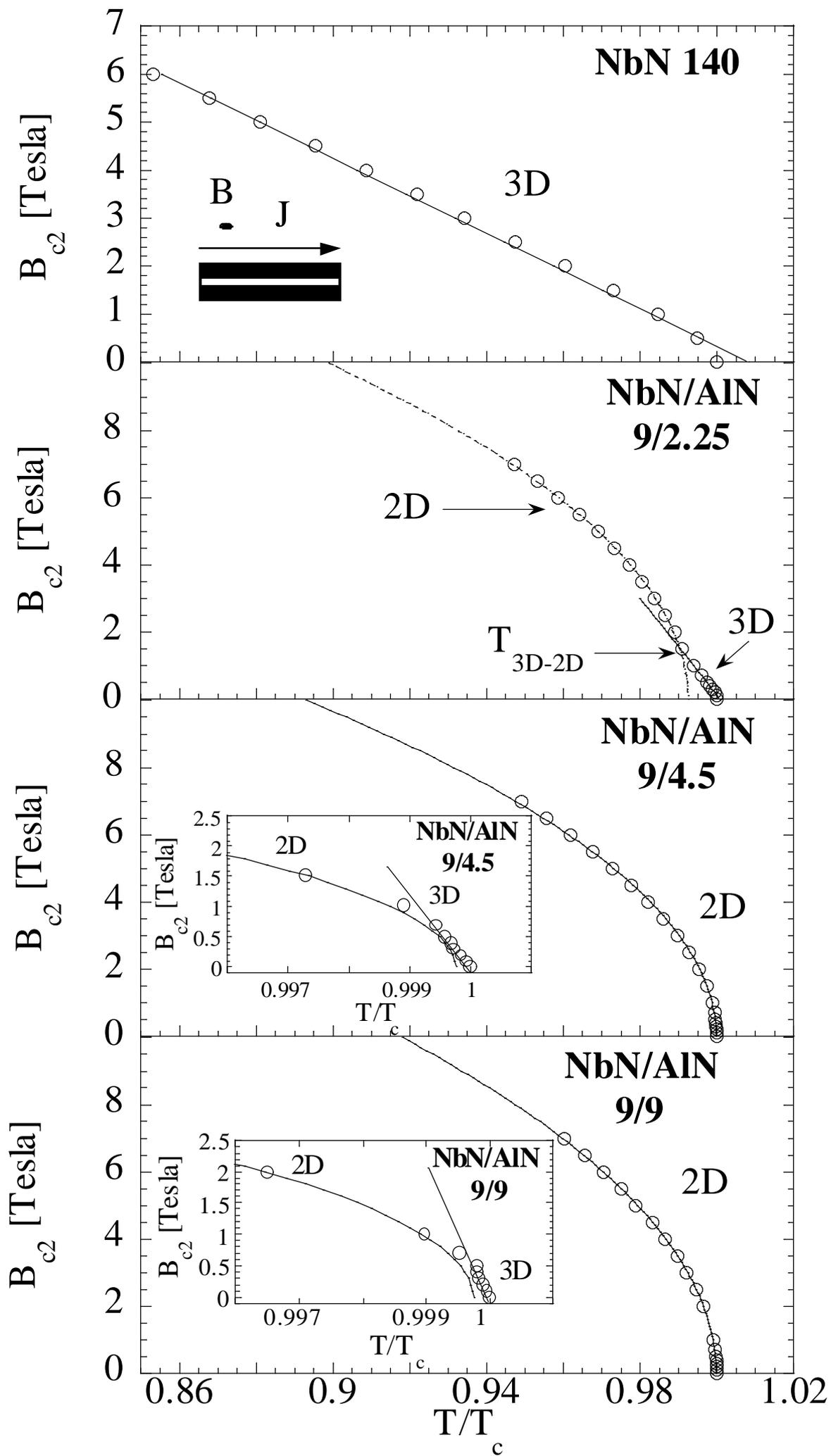

Fig. 3. E. S. Sadki *et al*

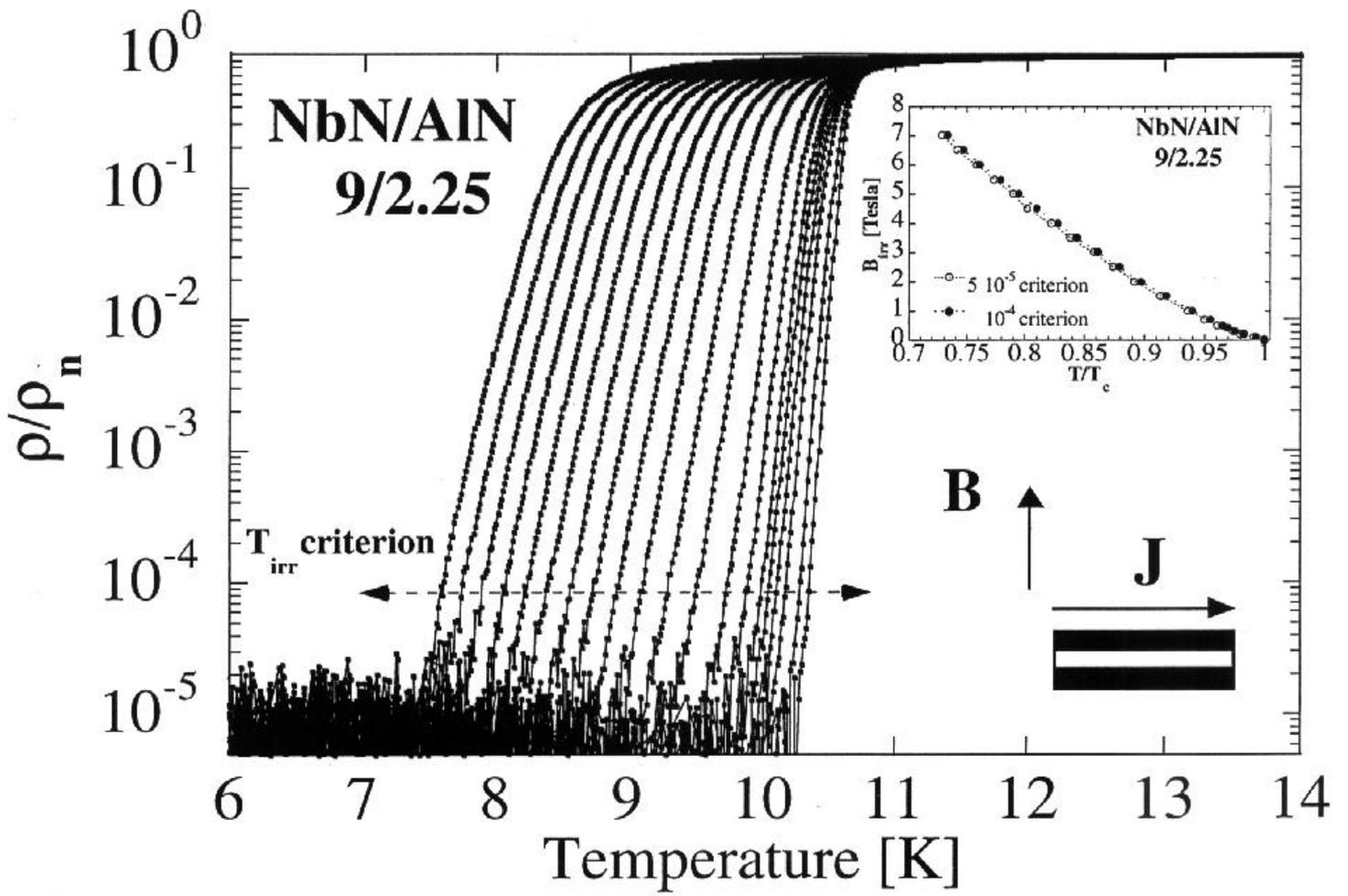

Fig. 4. E. S. Sadki *et al*

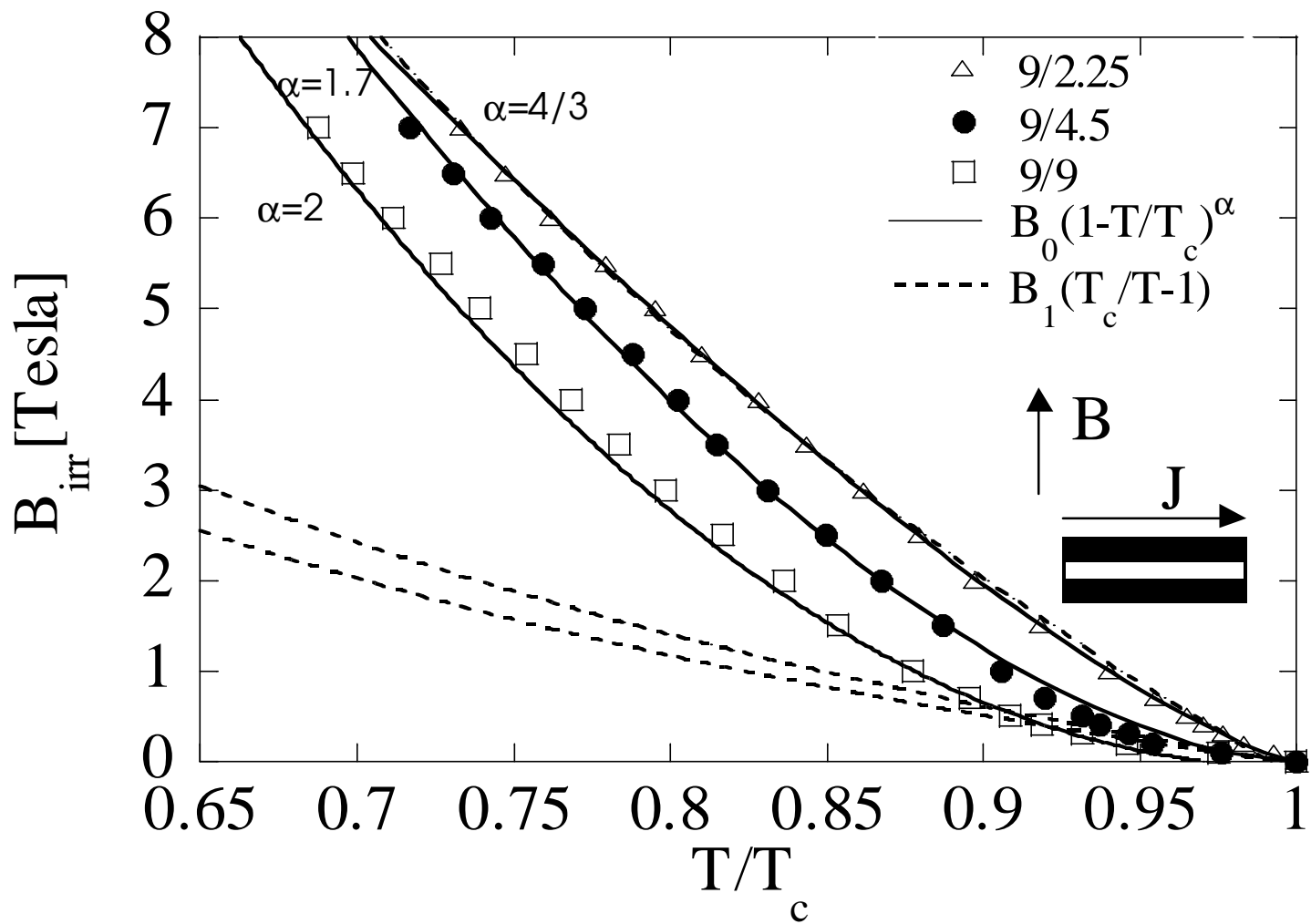

Fig. 5. E. S. Sadki *et al*

# Effects of interlayer coupling on the irreversibility lines of NbN/AlN superconducting multilayers

E. S. Sadki [1], Z. H. Barber [1,2], S. J. Lloyd [2],
M. G. Blamire [1,2] and A. M. Campbell [1]

[1] *Interdisciplinary Research Centre in Superconductivity, University of Cambridge, Cambridge CB3 0HE, United Kingdom*

[2] *Department of Materials Science and Metallurgy, University of Cambridge, Cambridge CB2 3QZ, United Kingdom*

We have studied the temperature dependence of the in-plane resistivity of NbN/AlN multilayer samples with varying insulating layer thickness in magnetic fields up to 7 Tesla parallel and perpendicular to the films. The upper critical field shows a crossover from 2D to 3D behavior in parallel fields. The irreversibility lines have the form $(1-T/T_c)^\alpha$, where $\alpha$ varies from 4/3 to 2 with increasing anisotropy. The results are consistent with simultaneous melting and decoupling transitions for low anisotropy sample, and with melting of decoupled pancakes in the superconducting layers for higher anisotropy samples.

The vortex matter in superconducting multilayers, and in high-$T_c$ superconductors (HTS) in particular, is subject to the competing roles of dimensionality, thermal fluctuations, and quenched disorder [1]. In these systems, the flux lines can be considered as a stack of 2D pancake vortices (PV) threading the superconducting layers ($CuO_2$ layers in HTS) and coupled via Josephson and magnetic interactions [2]. Thermal fluctuations can overcome the intralayer vortex order and the interlayer coupling between PV forming a flux line. The first phenomenon is referred to as melting of the vortex-solid into a vortex-liquid [3], and the second as decoupling [4]. Depending on the disorder of the system the vortex-solid can be vortex-lattice [5] or vortex-glass [6]. Experimentally, one of the most studied features in HTS is the so-called irreversibility line (IL) [7]. This line divides the *B-T* phase diagram into two parts: a magnetically irreversible, zero resistance state and a reversible state with dissipative electrical transport properties. In clean samples, IL is associated with a first-order vortex-lattice melting transition, illustrated by a sharp resistive kink [8] and a magnetization jump [9]. For highly disordered systems, it is associated with a second-order vortex glass melting transition [10]. Despite progress in this area, one of the remaining problems in HTS is to systematically study the effects of interlayer coupling (or anisotropy) on the vortex matter. Some groups have investigated the changes in IL as a function of oxygen doping in the $CuO_2$ layers [11]. However, this method affects simultaneously the anisotropy and the disorder, and therefore complicates the interpretation of the results. One way to address this problem is to use low-$T_c$ superconducting multilayers as a model system for HTS. The advantage of this method is that the layer thickness, and therefore the anisotropy, can be systematically and independently controlled. Several studies have been undertaken along these lines in systems such as Nb/Ge [12], MoGe/Ge [13], NbGe/Ge [14] and others [15].

In this Letter, we report studies of the in-plane properties of low-$T_c$ superconducting NbN/AlN multilayers. These systems are ideal for investigating the role of anisotropy in disordered layered superconductors because of their strong pinning properties. Consequently, in contrast to the previously studied weak pinning low-$T_c$ multilayers and HTS, the effects of thermal depinning [1], which can make the melting and decoupling transitions unobservable, are reduced. Using transport measurements in magnetic fields applied perpendicular and parallel to the layers, the effects of changing the insulating (AlN) layer thickness have been systematically investigated, and a monolithic NbN film was also studied for comparison. Unlike the NbN, the resistivity transition of NbN/AlN multilayers was found to broaden with increasing magnetic field. The upper critical fields for fields applied parallel to the layers show a dimensional crossover in the superconductivity. The irreversibility lines of the NbN/AlN multilayers and their evolution with increasing anisotropy are consistent with the theories of melting and decoupling in layered superconductors.

High quality NbN/AlN multilayers were deposited onto silica substrates at 360 $^0$ C using reactive dc-magnetron sputter deposition as described elsewhere [16]. The multilayers are referred to as *X/Y*, where *X* and *Y* represent the NbN and AlN layer thicknesses, respectively in nm. Each multilayer has the first NbN layer in contact with the substrate, followed by 10 AlN/NbN bilayers. The multilayer structures were characterized using X-ray diffraction and transmission electron microscopy (TEM). The multilayers consist of untextured, polycrystalline NbN, with predominantly amorphous AlN layers. The layer roughness is largely determined by the NbN grain size (which is on a similar scale to the NbN layer thickness) [16].

For the measurements reported here, films were patterned into 500 $\mu$m×50 $\mu$m tracks for the four-point configuration in order to measure the in-plane resistance. The contact pads were scratched before applying conductive silver paint to ensure that current was injected into all of the layers through the film thickness. Resistance was measured as a function of temperature in fields up to 7 Tesla applied parallel or perpendicular to the layers. The orientation of the sample was controlled to better than $0.1^0$. An ac current source was used at a frequency of 72 Hz and the voltage measured with lock-in amplifiers. The small currents used (1 $\mu$A) were chosen to optimize the sensitivity whilst ensuring that a very weak non-linearity in the resistivity does not affect our conclusions.

Figure 1 shows the resistivity transition for perpendicular applied fields for a 9/4.5 multilayer. The same set of data for a 140 nm thick NbN film is shown in the inset. The resistivity transition for the NbN film shifts to lower temperatures with increasing field with no marked change in form. On the other hand, the multilayer sample shows pronounced broadening with increasing field. The upper critical field, $B_{c2}$, is defined as the midpoint in the resistivity transition ($\rho/\rho_n = 0.5$), where $\rho_n$ is the normal state resistivity.

Figures 2 and 3 show the upper critical field ($B_{c2}$) for the 140 nm NbN sample together with that for 9/2.25, 9/4.5 and 9/9 multilayers, for perpendicular and parallel orientations of magnetic field, respectively. For perpendicular fields, the upper critical field for samples NbN 140 and 9/2.25, show a good linear fit, as expected from the Ginzburg-Landau (GL) theory for 3D superconductors close to $T_c$ [17]. For samples 9/4.5 and 9/9, the data could not be fitted to the linear prediction of GL theory, or to the Werthamer equation [18], because of the small curvature near $T_c$. This curvature has also been reported in reference [12] and seems to be universal for multilayers. However, taking an empirical power law fit proportional to $(1-T/T_c)^\alpha$ where $\alpha \approx 0.9$, gives us very reasonable values, which do not affect our later conclusions. The GL coherence length along the layers $\xi_{ab}(0)$, is defined by the equation $\xi_{ab}(0) = \sqrt{\Phi_0/2\pi B_{c2}(0)}$ [17], where $\Phi_0$ is the flux quantum and $B_{c2}(0)$ is obtained from the fitted curves. For parallel fields, the upper critical field for NbN 140 is linear as expected from GL theory. However, this linear dependence does not fit the data very close to $T_c$ (see fig 3). The reason is that for a thin film, in an applied magnetic field parallel to its surface, and very close to $T_c$, the coherence length becomes of the same order of magnitude as its thickness. Thus the film behaves like a 2D superconductor, and the linear dependence of $B_{c2}$ is lost [17]. For the multilayer samples, the behavior is more complex. At the lowest temperatures the curves are proportional to $(1-T/T_c)^{1/2}$, which is typical for 2D superconducting films [17]. This is explained by the Lawrence-Doniach (LD) model for layered superconductors [19]. At these temperatures, the coherence length across the layers $\xi_c(T)$, is much smaller than the layer spacing, and the layers behave like independent 2D superconducting films. However, approaching $T_c$, $\xi_c(T)$ becomes bigger than the interlayer spacing, and the whole multilayer behaves like a 3D superconductor leading to a linear dependence. The crossover temperature $T_{3D-2D}$ is obtained from the intersection of the two regimes. In the 3D regime ($T > T_{3D-2D}$) the upper critical field is defined by GL theory for anisotropic superconductors, i.e. $B_{c2} = \Phi_0/2\pi\xi_{ab}\xi_c$ [17]. Using the values of $B_{c2}(0)$ obtained from the 3D curve fits, and the values of $\xi_{ab}(0)$, $\xi_c(0)$ is obtained. Finally, the anisotropy ratio $\Gamma = \xi_{ab}/\xi_c$ is calculated. The results are shown in Table I.

The irreversibility temperature $T_{irr}(B)$ for each magnetic field was defined close to the limit of the sensitivity of the measured resistivity on a logarithmic scale at $\rho/\rho_n = 10^{-4}$ (see fig. 4). To confirm that our general conclusions are not dependent upon the criterion used, we plotted the IL (inset of figure 4) obtained from a lower criterion $\rho/\rho_n = 5\times10^{-5}$ together with the first one. It is clear that the two graphs coincide. Figure 5 shows the IL for the multilayer samples. The data are fitted to the equations $B_0(1-T/T_c)^\alpha$ and $B_1(T_c/T-1)$. Both $B_0$ and $B_1$ are field and temperature independent. The physical meaning of these fits is discussed below. For sample 9/2.25, both curves can be fitted very well for the whole range of data points available, with $B_0 = 39\,\text{T}$, $\alpha = 4/3$ and $B_1 = 19.8\,\text{T}$. For samples 9/4.5 and 9/9, the best fits for the first equation give $B_0 = 59.5\,\text{T}$ and 70.9, and $\alpha = 1.7$ and 2, respectively. However, for both samples the second equation could only be fitted very close to $T_c$, with $B_1 = 5.7$ and 4.8 T, for 9/4.5 and 9/9, respectively.

Thermal fluctuations are governed by the Ginzburg number $Gi$, given by $Gi = 1/2\left(2\pi\mu_0 k_B T_c \lambda_{ab}^2(0)\Gamma/\Phi_0^2 \xi_{ab}(0)\right)^2$, where $\lambda_{ab}$ is the inplane penetration depth [1]. In HTS, $Gi \approx 10^{-2}$ for YBCO and $Gi \approx 1$ for BSCCO [1]. This number is much smaller in bulk low-$T_c$ materials. For example, $Gi \approx 4\times 10^{-4}$ in sample NbN 140, where we have used the value $\lambda_{ab}(0) = 0.5\,\mu\text{m}$ from reference [20]. However, since $Gi$ is proportional to $\Gamma^2$, thermal fluctuations are dramatically enhanced in the multilayers. For our samples, $Gi$ is in the range between 4 $10^{-3}$ and $5\times 10^{-1}$. These values are of the same order of magnitude to those of HTS. These large thermal fluctuations can either lead to melting or depinning. Melting occurs when the thermal energy $k_B T$ is of the same order of magnitude as the vortex-vortex energy, and depinning occurs when $k_B T \approx U_p$, where $U_p$ is the pinning energy. However, one of the most difficult problems in type II superconductors is to evaluate $U_p$ directly [21]. The collective pinning theory [22], which suggests that the vortices are pinned collectively in a correlation volume $V_c$, was used successfully to describe the results in low-$T_c$ NbGe [23]. However, it was found to be difficult to apply it to stronger pinning systems [24]. To avoid this problem, we estimate the level of pinning strength from the dimensionless ratio $J_c/J_0$, where $J_c$ is the critical current density, and $J_0 = \Phi_0/(3\sqrt{3}\pi\lambda_{ab}^2(0)\xi_{ab}(0)\mu_0)$ is the depairing current [1]. For example, $J_c/J_0 \approx 10^{-3}-10^{-2}$ in HTS, $4\times10^{-6}$ in low pinning NbGe (calculated using $J_c = 10\,\text{A/cm}^2$ from [25]), and $10^{-2}$ in moderate pinning MoGe [13]. It is well known that NbN is a strong pinning system due to its granular structure [20], [26]. For this system $J_c/J_0$ is calculated to be about $7\times10^{-1}$ with $J_c \approx 10^6\,\text{A/cm}^2$ from reference [26]. In the range of temperatures where there is a strong evidence of melting in HTS, the thermal energy is about 10 times bigger than it is in our experiments. Furthermore, the pinning strength in NbN is about one order of magnitude bigger than in HTS. Also melting has been reported in both NbGe thin films [25] and NbGe/Ge multilayers [14], where the pinning strength is much lower. All this suggests that it is unlikely for thermal depinning to occur at lower temperatures than melting in our samples.

Before discussing the melting and decoupling transitions, it is important to define the dimensionality of the vortices. In the case where the magnetic field is applied perpendicular to the multilayers, the vortex system is considered to be 3D if the PV in adjacent layers are strongly correlated so as to define a continuous vortex line [2]. It is 2D if the PV in adjacent layers move independently. It is expected that the decoupling occurs when the shear energy starts to exceed the tilt energy [4]. This estimate gives a decoupling field $B_{2D} \approx 4\Phi_0/\Gamma^2 s^2$, where $s$ is the spacing between the superconducting layers [4]. $B_{2D}$ is between 117 and 230 T for YBCO ( $s = 1.5$ nm and $\Gamma = 5-7$ ), and between 0.9 and 1.5 T for BSCCO ( $s = 1.2$ and $\Gamma = 50-200$ ) [1]. For the multilayers, we obtain 119, 0.6 and 0.06 T, for 9/2.25, 9/4.5 and 9/9, respectively. We conclude that for sample 9/2.25 the applied magnetic fields in our experiments are not high enough to decouple the PV, so that if there is a decoupling transition it should be generated by thermal fluctuations only. However, for samples 9/4.5 and 9/9, it is clear that PV are decoupled for most of the range of applied fields. From a simple Lindemann criterion, the melting equations are of the form $B_m(T) = B_m(0)(1 - T/T_c)^\alpha$, where $B_m(0) = c_L^4 \Phi_0^5 / 12\pi k_B^2 T_c^2 \Gamma^2 \mu_0^2 \lambda_{ab}^4(0)$, $0.1 \leq c_L \leq 0.2$ and $\alpha \leq 2$ [3]. By comparing our experimental results and this equation, we obtain $c_L = 0.18$, 0.5 and 0.7, for 9/2.25, 9/4.5 and 9/9, respectively. It is clear that this theory gives the right range of values for sample 9/2.25. For the other two samples, the values of $c_L$ are outside the predicted range. This can be explained by the fact that the PV are decoupled for almost the whole range of temperatures and fields studied here, and that the previous melting equation was originally derived in the case of 3D vortices. However, because of the strong pinning in our system, we believe that the IL in 9/4.5 and 9/9 represents a 2D melting of the PV in the superconducting layers. The decoupling field is given by $B_{dc}(T) = B_1(T_c/T - 1)$, where $B_1 = \Phi_0^3 / 4\pi^2 e \mu_0 k_B T_c s \lambda_{ab}^2(0) \Gamma^2$ and $e = 2.718$ [4]. This gives $B_1 = 57.6\,\mathrm{T}$ for sample 9/2.25. This value is bigger than the measured one, although it is of the same order of magnitude. In conclusion both the melting and decoupling line fit for the whole range of data for sample 9/2.25, suggesting that they occur simultaneously within the range of field and temperature studied here. The difficulty found in fitting the decoupling equation (except very close to $T_c$) in samples 9/4.5 and 9/9 is justified here since the PV are already decoupled due to the effect of the applied magnetic field. However, we cannot make a quantitative estimate of the nature of the transition very close to $T_c$, i.e. distinguish between melting and simultaneous melting and decoupling. Finally, we note that our resistivity data, in contrast to weak pinning low-$T_c$ multilayers and clean HTSC, do not show the characteristic kink or jump corresponding to a first order melting transition. This suggests that the reported transitions in this paper correspond to a continuous (second order) vortex-glass melting due to the presence of strong pinning in our systems.

In summary, we have studied the temperature dependence of the in-plane resistance of a NbN film and three NbN/AlN multilayer samples with varying insulating layer thickness in magnetic fields parallel and perpendicular to the films. The upper critical field shows a crossover from 2D to 3D behavior in parallel fields. The IL in the low anisotropy sample is consistent with simultaneous melting and decoupling transitions. For higher anisotropy samples it is consistent with melting of decoupled pancakes in the superconducting layers.

We would like to thank M. McElresh, M. J. W. Dodgson, N. H. Babu, S. F. W. R. Rycroft and A. D. Bradley for useful discussions. One of us (ESS) would like to thank the Algerian Government and Trinity College, Cambridge, for financial assistance during his Ph.D.

**Figure captions**

FIG. 1. Normalized resistivity versus temperature for sample 9/4.5 for magnetic fields perpendicular to the layers from 0 to 7 Tesla. The inset shows the same data for single layer NbN (140 nm).

FIG. 2. Upper critical field perpendicular to the layers for NbN 140, 9/2.25, 9/4.5 and 9/9 samples (see text).

FIG. 3. Upper critical field parallel to the layers for NbN 140, 9/2.25, 9/4.5 and 9/9 samples, showing the 2D-3D transition for 9/2.25. The insets show this transition (close to $T_c$) for 9/4.5 and 9/9 samples (see text).

FIG. 4. Normalized resistivity versus temperature graph showing the definition of $T_{irr}$ at fixed $B$ for the 9/2.25 sample. The inset shows irreversibility $T_{irr}(B)$ defined at different criteria as a function of the reduced temperature $T/T_c$.

FIG. 5. Irreversibility lines of 9/2.25, 9/4.5 and 9/9 samples for perpendicular magnetic fields. The data is fitted to the melting (continuous line) and decoupling (dashed) equations (see text).